\documentclass{mn2e}
\usepackage{graphicx}
\title[GAMA: Mass-to-light ratios]{Galaxy And Mass Assembly (GAMA): The Consistency of GAMA and WISE Derived Mass-to-Light Ratios}
\author[T. Kettlety et al.]
{T. Kettlety$^{1,2}$, J. Hesling$^1$, S. Phillipps$^{1}$\thanks{s.phillipps@bristol.ac.uk}, M.N. Bremer$^{1}$, 
M.E. Cluver$^{3}$, 
\and
E.N. Taylor$^4$, J. Bland-Hawthorn$^5$, S. Brough$^6$, R. De Propris$^7$, S.P. Driver$^{8,9}$, 
\and B.W. Holwerda$^{10}$, L.S. Kelvin$^{11}$, W. Sutherland$^{12}$, A.H. Wright$^{13}$\\
$^{1}$Astrophysics Group, H.H. Wills Physics Laboratory, University of Bristol, Tyndall Avenue,  Bristol BS8 1TL, UK\\
$^2$School of Earth Sciences, University of Bristol, Queens Road, Bristol BS8 1RJ, UK\\
$^3$Department of Physics and Astronomy, University of the Western Cape, Bellville 7535, South Africa\\
$^4$Centre for Astrophysics and Supercomputing, Swinburne University of Technology, Hawthorn, Vic 3122, Australia\\
$^5$Sydney Institute for Astronomy, School of Physics A28, University of Sydney, NSW 2006, Australia\\
$^6$School of Physics, University of New South Wales, NSW 2052, Australia\\
$^7$Finnish Centre for Astronomy with ESO, University of Turku, Vaisalantie 20, Piikkio, Finland\\
$^8$International Centre for Radio Astronomy Research, University of Western Australia, Crawley, WA 6009, Australia\\
$^9$School of Physics \& Astronomy, University of St Andrews, North Haugh, St Andrews KY16 9SS, UK\\
$^{10}$Department of Physics and Astronomy, University of Louisville, Louisville KY 40292, USA\\
$^{11}$Astrophysics Research Institute, Liverpool John Moores University, 146 Brownlow Hill, Liverpool L3 5RF, UK\\
$^{12}$School of Physics and Astronomy, Queen Mary University of London, Mile End Road, London E1 4NS, UK\\
$^{13}$Argelander-Institut f\"{u}r Astronomie, Universit\"{a}t Bonn, Auf dem H\"{u}gel 71, D-53121 Bonn, Germany
}

\begin{document}

\date{Second Draft. Accepted YYYY Month DD. Received YYYY Month DD; in original form YYYY Month DD}

\pagerange{\pageref{firstpage}--\pageref{lastpage}} \pubyear{YYYY}

\maketitle

\label{firstpage}

\begin{abstract}
Recent work has suggested that mid-IR wavelengths are optimal for estimating the mass-to-light ratios of stellar populations and hence the stellar masses of galaxies. We compare stellar masses deduced from spectral energy distribution (SED) models, fitted to multi-wavelength optical-NIR photometry, to luminosities derived from {\it WISE} photometry in the $W1$ and $W2$ bands at 3.6 and 4.5$\mu$m for non-star forming galaxies. The SED derived masses for a carefully selected sample of low redshift ($z \le 0.15$) passive galaxies agree with the prediction from stellar population synthesis models that $M_*/L_{W1} \simeq 0.6$ for all such galaxies, independent of other stellar population parameters. The small scatter between masses predicted from the optical SED and from the {\it WISE} measurements implies that random errors (as opposed to systematic ones such as the use of different IMFs) are smaller than previous, deliberately conservative, estimates for the SED fits. This test is subtly different from simultaneously fitting at a wide range of optical and mid-IR wavelengths, which may just generate a compromise fit: we are directly checking that the best fit model to the optical data generates an SED whose $M_*/L_{W1}$ is also consistent with separate mid-IR data. We confirm that for passive low redshift galaxies a fixed $M_*/L_{W1} = 0.65$ can generate masses at least as accurate as those obtained from more complex methods.
 Going beyond the mean value, in agreement with expectations from the models, we see a modest change in  $M_*/L_{W1}$ with SED fitted stellar population age but an insignificant one with metallicity.
\end{abstract}


\section{Introduction}
Over the past three decades, increasingly large samples of galaxies have been surveyed in terms of their photometric and spectroscopic (primarily redshift) properties (e.g. 2dFGRS, Colless et al. 2001; SDSS, York et al. 2000, Alam et al. 2015; GAMA, Driver et al. 2009). One of the key parameters which one would like to derive from such surveys is the stellar mass of the galaxies (e.g. Kauffmann et al. 2003a and Mendel et al. 2014 for SDSS; Taylor et al. 2011 for GAMA). This is clearly a factor in many areas, for instance the stellar mass function itself (e.g. Baldry et al. 2012, Thanjavur et al. 2016, Wright et al. 2017), the relationship between stellar and halo mass (e.g. Moster et al. 2010), the determination of the role of dark matter in generating rotation curves (e.g. Martinsson et al. 2013), the exploration of correlations between other stellar population parameters such as age or metallicity with stellar mass (e.g. Gallazzi et al. 2005), or studies of the star forming main sequence and specific star formation rates (e.g. Noeske et al. 2007, Davies et al. 2016), amongst others. Indeed, virtually all global quantities used to describe galaxies correlate with stellar mass (e.g. Kauffmann et al. 2003b, Brough et al. 2017).

Numerous related techniques have been explored for determining the stellar masses of galaxies, though all are based ultimately on stellar population synthesis (SPS) models to generate corresponding stellar mass-to-light ratios ($M_*/L$). The two main classes of techniques derive $M_*/L$ (in some particular observed band) either from a straightforward use of individual or multiple broadband colours (e.g. Bell et al. 2003, Gallazzi \& Bell 2009, Taylor et al. 2011) or from inverting fits to the whole spectral energy distribution (SED) across some wavelength range to obtain the stellar population parameters (e.g. Tojeiro et al. 2007, da Cunha et al. 2008; see Walcher et al. 2011 and Conroy 2013 for reviews).

If we are to have confidence in the results (to a reasonable level of accuracy), then implicitly we first need to believe the SPS model predictions, or at least their self-consistency, for as wide a range of data as possible. A fairly direct test of the models' capabilities in this regard is to check whether a given model generates the same mass (and preferably other ancilliary stellar population parameters such as age and metallicity) when used in different ways and/or with different data.

Recently there has been considerable interest in the use of mid-IR wavelengths for determining masses, since the mid-IR $M_*/L$ is relatively insensitive to other factors, i.e. is fairly constant for different stellar populations, especially in the absence of ongoing star formation (e.g. Wen et al. 2013, R\"{o}ck et al. 2015). In particular, Meidt et al. (2014) have suggested that using {\it Spitzer} or {\it WISE} measurements around $4\mu$m, e.g. the {\it WISE} $W1$ and $W2$ bands, $M_*/L_{W1}$ for non-star forming galaxies can be taken to have an expectation value of 0.6 (in the usual solar units) with only a relatively small scatter. Norris et al. (2014) have expanded on this and discussed the joint dependence of $M_*/L_{W1}$ and the ($W1 - W2$) colour on the stellar population age and metallicity. In the models they use, $M_*/L_{W1}$ is predicted to vary systematically with age and to a lesser extent with metallicity, while the mid-IR colour varies almost purely with metallicity (see also Norris et al. 2016). 

In the present paper, we investigate comparisons of the mid-IR route to stellar masses with respect to that of optical to near-IR SED fitting from multi-band photometry, as exemplified by the work of Taylor et al. (2011; henceforth T11). The work is based on the photometry accumulated for the GAMA project (see Hill et al. 2011), to which matched {\it WISE} data have been added by Cluver et al. (2014). We can therefore make direct comparisons of the two methods of determining masses on a galaxy-by-galaxy basis for a large sample of low redshift galaxies of a range of luminosities. 

Note that this is subtly different from the experiment of simultaneously fitting the SED across a wide range of wavelengths (e.g. Chang et al. 2015). We are specifically checking that the best fitting SED model to the optical data (in this case) for a given galaxy implies a stellar population which is also consistent with the fit to a separate set of photometric data (here mid-IR) for the same galaxy (see McGaugh \& Schombert 2014 for a related approach). 

Adding further wavebands will always give some best fit  (e.g. Poudel et al. 2016), but this could be a compromise between matching in the different wavelength regimes. Indeed the resulting fits may be worse than when using, say, the optical data on its own (see, e.g., the extensive discussion in T11 of the merits, or otherwise, of adding the then available near-IR data to the optical).

In this paper, we will be dealing with generally old stellar populations, so as the GAMA database supplies a number of indicators of recent star formation (e.g. Davies et al. 2016, Gunawardhana et al. 2013), we can utilise these in order to refine our samples. Other {\it WISE} colours also provide measures of star formation and AGN activity (Jarrett et al. 2013).

We structure the paper as follows. Section 2 reviews the available data, both optical and mid-IR, and its use to derive stellar masses, and Section 3 presents our comparison of the mass estimates. Section 4 summarises and discusses the relevance of the results for general mass estimation. 

Where required we use a standard concordance cosmology, i.e. $H_0 = 70$~km~s$^{-1}$Mpc$^{-1}$, $\Omega_m = 0.3$, $\Omega_{\Lambda} = 0.7$, as in T11 and in the GAMA catalogues. AB magnitudes are used for GAMA's optical data, while the {\it WISE} magnitudes are in the Vega system (see Jarrett et al. 2011). We take the Sun's absolute magnitude in the $W1$ band to be 3.24 (as in Cluver et al. 2014).

\section{Data and Models}

\subsection{Optical Data and $M_*/L$}

The GAMA (Galaxy And Mass Assembly) survey is in essence a redshift survey of five regions of the sky, with total area 286 square degrees, down to a magnitude limit of $r=19.8$, with observations made with the AAOmega spectrograph on the Anglo-Australian Telescope; see Liske et al. (2015) for a recent summary of GAMA Data Release 2. The survey was based on SDSS photometry that has been reprocessed and homogenized to give improved magnitudes (Hill et al. 2011, Kelvin et al. 2012). Note that both aperture/isophotal magnitudes and (asymptotic) total magnitudes are available, and it is important to distinguish between these as appropriate (using the {\em fluxscale} parameter in the GAMA stellar mass catalogue, as derived from radial profile fits in Kelvin et al. 2012). 
Besides the magnitudes and redshifts, derived properties such as distances and luminosities, are also provided, as described in Liske et al. (2015).\footnote{See also www.gama-survey.org/dr2/.} Of particular relevance here are the catalogued stellar population and dust extinction parameters obtained from fits to the SEDs (T11), and the inferred masses and $M_*/L$ ratios. In addition, GAMA provides spectral line measurements (Hopkins et al. 2013) and, where appropriate, the derived star formation rate (Gunawardhana et al. 2013, Davies et al. 2016).

The GAMA catalogued masses\footnote{We specifically use the GAMA catalogue StellarMassesv18, an updated version of that discussed in T11.} are derived, simultaneously with all other relevant stellar population parameters, from matched aperture photometry 
in the five optical SDSS bands $ugriz$, plus VISTA-VIKING $ZYJHK$ data (Edge et al. 2013), weighted so that only the restframe wavelength range 3000-10000\AA, i.e. restframe $u$ to $Y$ (henceforth `optical'), is actually utilised.\footnote{It was originally found that adding near-IR magnitudes from UKIDSS hindered, rather than helped, the fitting process, so these were not used by T11. The near-IR magnitudes from VISTA-VIKING imaging, on the other hand, {\em are} found to be consistent with extrapolating the SED fits previously made; see Driver et al. (2016) for examples.} Full details of the modelling and fitting process are given in T11, but we can summarise by noting that SED templates are used which are based on the simple stellar population (SSP) evolutionary models of Bruzual \& Charlot (2003; henceforth BC03), for a range of stellar metallicities $Z$, and the stellar initial mass function (IMF) of Chabrier (2003). In the fitted composite stellar populations (CSP), these are weighted according to a star formation history (SFH) which begins a time $t$ before the epoch of observation (the `age') and has an exponential fall-off with time constant $\tau$. For simplicity, dust extinction is assumed to be uniform (i.e. a foreground screen in front of the stars) and follow the Calzetti et al. (2000) extinction law, quantified via $E(B-V)$. Besides the best fitting SED, the process therefore also returns values for $t$, $\tau$, $Z$ and $E(B-V)$.

The best-fitting stellar population model and rest-frame, dust-corrected (i.e. intrinsic stellar) SED can then be used to calculate $M_*/L$ at any wavelength and hence the stellar mass from the known luminosities. T11 also compared their derived $M_*/L$ values to those calculated from a simple conversion between intrinsic stellar $(g-i)$ colour and $M_*/L_i$ and found consistency between them. They noted that the full (optical) SED fitting did, as would be hoped, add some extra information and constraints (see also Bell \& de Jong 2001, Zibetti et al. 2009). However, the empirical relation between $(g-i)$  and $M_*/L_i$ has the major advantage of simplicity, especially for comparing masses across surveys. This comes with little significant cost in terms of accuracy because of (in this case helpful) degeneracies between age, metallicity and extinction, which give rise to closely similar values of $M_*/L_i$ at a given colour, even if the population parameters are not individually well determined.

\subsection{WISE Data and $M_*/L$}

Cluver et al. (2014) discussed matching {\it WISE} data in the four {\it WISE} bands ($W1$ to $W4$ at 3.6, 4.5, 12 and 22$\mu$m) to the corresponding GAMA data. A GAMA catalogue provides matched {\it WISE} magnitudes from the {\it ALLWISE} catalogue, specifically what are referred to as ``recommended''  magnitudes in Cluver et al. (2014). These are based on the optimum treatment of individual sources depending on the signal-to-noise ratio and degree of resolution of each image, with appropriate aperture corrections (see Cluver et al. 2014 for details).  We then use these luminosities together with the T11 stellar masses to calculate $M_*/L_{W1}$. Equivalently, we could derive stellar masses from the $W1$ luminosities and a predicted $M_*/L_{W1}$ and compare to T11 masses (see below). Note that the WISE magnitudes are pseudo-total ones, so we use the version of GAMA masses which include the {\it fluxscale} correction. We retain only those galaxies for which this correction is less than $0.5\:$dex. Further GAMA catalogues now contain {\it WISE} fluxes and stellar masses derived using the {\sc LAMBDAR} software (see Wright et al. 2016), but as these use the {\it WISE} data as well as the optical data in determining the population fits, they are not appropriate for our present purpose.

Meidt et al. (2014) found that ($W1-W2$) colour should be a good indicator of $M_*/L_{W1}$ which minimises many uncertainties such as the effects of dust and details of post-main sequence stellar evolution. Indeed, they found that for a wide range of simple (or generally old) stellar populations the global assumption of $M_*/L_{W1} = 0.6$ should deliver stellar masses to a comparable accuracy to more complex methods, even without allowing for ($W1-W2$). Norris et al. (2014) extended this discussion and (albeit with a slightly different SSP model) determined where different age and metallicity stellar populations should lie in the plane of ($W1-W2$) colour and $M_*/L_{W1}$. They also presented observed results for samples of globular clusters and a small number of early type galaxies. R\"{o}ck et al. (2015) also find that, at the ages and metallicities of interest here (see below), $M_*/L_{W1}$ should be close to 0.6 over a range of inputs for the SPS modelling.

\subsection{The Galaxy Sample and Derived Parameters}
With the above in mind, our sample was selected from the initial matched {\it WISE}-GAMA sample according to a number of criteria. Firstly, we chose galaxies out to a redshift limit of $z=0.15$. This has a number of advantages, for instance we remove many of the faint objects with larger magnitude (and therefore SED) errors (T11), and obviously the spectral shift itself is fairly small, so that uncertain k-corrections in the mid-IR are minimised. (No k-correction is needed in the optical, since restframe photometry is derived in the course of the SPS fits used to obtain the stellar masses). We also remove any sources at $z<0.003$ (generally contaminating stars).

In order to restrict comparisons to, as far as possible, old stellar populations, where we might hope the models are most secure, we have attempted to remove AGN and star-forming galaxies. Selecting only passive galaxies should provide the cleanest test of the SPS models, minimising any effects from the star formation history. We first remove galaxies with measurable emission lines in the GAMA data (Hopkins et al. 2013). Further, we assume that {\it WISE} $(W1-W2) > 0.8$ indicates the presence of an AGN (Stern et al. 2012), while {\it WISE} $(W2-W3) > 1.5$ implies star formation (Jarrett et al. 2013, Cluver et al. 2014). This leaves a total of 718, assumed passive, non-AGN, galaxies. The {\it WISE} ($W2-W3$) cut removes a significant number of objects that could otherwise have been classed as non-star forming because their emission lines were too weak to measure accurately. Theoretically, we would expect galaxies to be passive and blue in ($W2-W3$) or star forming and red, but measurement errors can obviously shuffle objects across the boundaries. Our approach of requiring {\it both} lack of significant emission lines and blue mid-IR colours should provide the most secure sample of passive galaxies, as required for the present work.\footnote{Here, as elsewhere in this work, we use TOPCAT (Taylor M.B. 2005) to manipulate the various GAMA catalogues.} 

In order to calculate consistent rest frame $W1$ band luminosities across our sample, we require a suitable k-correction for the higher redshift galaxies. As there is currently no standardised k-correction for these wavelengths in the GAMA-{\it WISE} catalogues, we calculated a simple k-correction based on an assumed power law mid-IR SED, $F_{\lambda} \propto \lambda^n$, viz. $k_{W1} = -2.5(1+n)$~log$(1+z)$. We can determine the effective slope $n$ for a source with zero {\it WISE} $(W1-W2)$ colour from the ratio $F_{W1}/F_{W2}$ as given by the zero points of the {\it WISE} magnitude scale (Jarrett et al. 2011). This gives $n = -3.85$ (unsurprisingly close to the $n=-4$ of the Rayleigh-Jeans tail of a blackbody curve, see e.g. Brown et al. 2014) and hence $k_{W1} = -7.1$~log$(1+z)$. For our selected redshift range, this gives a maximum correction of $-0.43$ magnitudes (mid-IR k-corrections are negative since $F_{\lambda}$ is decreasing with $\lambda$). This is quite similar to the correction from Huang et al. (2007), as used by e.g. Neil et al. (2016), which is linear in $z$ and reaches $-0.34$ magnitudes at $z=0.15$.

Typically, our galaxies have colours within 0.1 magnitudes of the zero colour assumption and this will change the effective $n$ by only $\pm 0.3$. This in turn will change the k-correction, even for our most distant objects, by only 0.045 magnitudes, leading to changes of less than 5\% in derived luminosities. As this is smaller than other likely errors, we do not attempt to use a colour-dependent k-correction.  A small additional uncertainty in the k-correction could potentially arise from the presence of $3.3\mu$m PAH emission (e.g. Querejeta et al. 2015), though this emission should be very small for the selected type of galaxy and in the redshift range chosen it would always be contained within the $W1$ bandpass.

For the colours themselves, we make an empirical correction to allow for an observed trend of the measured $(W1-W2)$ becoming redder with redshift (c.f. Yan et al. 2013). This should remove any differential k-correction (for a pure power law, the k-correction would be the same in each band) as well as any evolutionary effects or selection biases. This linear correction is 0.18 magnitudes at our maximum $z = 0.15$, relative to the colour at $z=0$.

\subsection{SPS Predictions for the Mid-IR}
As noted, the optical SED fitting of T11 assumes a BC03 stellar evolution model, a Chabrier IMF and a screen of absorbing dust modelled as in Calzetti et al (2000). Meidt et al. (2014) modelled the mid-IR photometry also via BC03 models with a Chabrier IMF, but assumed minimal dust absorption at these wavelengths (though they correct their data for potential PAH and hot dust {\em emission}; see Querejeta et al. 2015 for a detailed discussion of this emission in star-forming galaxies). They argue that the model predictions at these wavelengths are much less susceptible to the uncertainties due to the details of the SPS treatment, with the 3.6$\mu$m emission dominated by old stars on the red giant branch (e.g. Spitler et al. 2008, da Cunha et al. 2008, Peletier et al. 2012, R\"ock et al. 2015) and that there is much less variation in mass-to-light ratio due to age/SFH and metallicity than seen at shorter wavelengths. They also argue that the treatment of the contribution of AGB stars in the original BC03 models better matches observations than do those in either the more recent 2007 version of BC03, or the alternative prescription in Maraston (2005). McGaugh \& Schombert (2014) have come to similar conclusions.

Meidt et al. (2014) note that $M_*/L$ at 3.6$\mu$m should, nevertheless, increase with the average age of the stellar population and decrease with increasing metallicity (the opposite of optical $M_*/L$ ratios), with an overall combined possible range of order 0.3 dex (factor 2) for the extremes of a very wide range of old metal-poor systems to young metal-rich systems. Allowing all of these possible populations would induce a (1 $\sigma$) scatter of around 0.11 dex (30\%). Incorporating constraints from the mid-IR colours (which depend primarily on metallicity, rather than age), they conclude that the uncertainty in $M_*/L_{W1}$ can be reduced to about 0.07 dex ($\simeq$ 20\%). Furthermore, even with no other observed constraints on age or metallicity, they suggest that both old, metal-rich and younger, metal-poor populations - the most likely combinations given the known age-metallicity relation (AMR) for elliptical galaxies (Gallazzi et al. 2005) - should actually have the same $M_*/L_{W1} =0.6$ to within 0.06 dex ($\simeq$ 15\%). 

However, Meidt et al. (2014) also note that models such as BC03 do not predict the correct absolute 3.6 to 4.5$\mu$m {\it Spitzer} colours for observed galaxies, probably due to the lack of treatment of the CO absorption feature in the 4.5$\mu$m band causing the models to be too red. Meidt et al. (2014) therefore determined an empirical shift (based on the observed near-IR and mid-IR colours of giant stars) of the BC03 models to match observed galaxy mid-IR colours.

Following this, Norris et al. (2014) used instead the Bressan et al. (2012) SSP model (which attempts to correct for the effect of the CO feature) to determine $M_*/L_{W1}$ and the mid-IR {\it WISE} colours for a variety of SSPs. They confirm that $M_*/L_{W1}$ has some dependence on both age and metallicity for intermediate to old stellar populations while ($W1-W2$) is insensitive to age but becomes slightly bluer with increasing metallicity (again attributing this to the CO feature), and demonstrate that their model matches the trends in observed data.  

Specifically, the Norris et al. (2014) results (e.g. their figure 5) imply that between ages of 3 (or 5) and 10~Gyr, SSPs (at given metallicity) should increase their $M_*/L_{W1}$ by approximately 0.3 (or 0.18) dex. The range of ages used by Meidt et al. (2014) - based on similar declining SFHs to those used by T11 - suggest a similar variation $\simeq 0.2$ dex. 
In terms of metallicity, Norris et al. find that at fixed age, the model log($M_*/L_{W1}$) increases by approximately 0.04 between $Z_{\odot}$ and 0.1 $Z_{\odot}$while Meidt et al. predict a compatible change of $\sim 0.02$ dex between 1 and 0.2 $Z_{\odot}$.
(See also R\"{o}ck et al. (2015) for a similar treatment and results using a variety of SPS model ingredients). 

The range and scatter in $M_*/L_{W1}$ for real galaxies can be empirically quantified for our passive sample, and we carry out this exercise below. Clearly the observed scatter will be a combination of real physical (systematic) variations, arising from the range of stellar populations which are included, as discussed above, stochastic galaxy-to-galaxy variations and observational uncertainties.

\section{Results}
\subsection{Masses}

We first use the GAMA optically-derived stellar masses and the  $W1$ luminosities calculated as in the previous section to determine the  mass-to-light ratios $M_*/L_{W1}$ for our sample of passive galaxies. In Figure 1 (top panel) we follow Norris et al. (2014) and plot these against the redshift corrected $(W1-W2)$ colours. Note that the majority of points lie within $-0.3 \leq$ log$(M_*/L_{W1}) \leq -0.1$ and $-0.15 \leq (W1-W2) \leq +0.05$. As there is no evidence for a trend in $M_*/L_{W1}$ with colour, we can simply fit a Gaussian to the distribution of log$(M_*/L_{W1})$ values for the whole sample (Figure 1 lower panel), obtaining a mean of $-0.19$ and standard deviation 0.05, corresponding to $M_*/L_{W1} = 0.65 \pm 0.07$. The first thing to note, therefore, is that globally the data (i.e. the GAMA optically derived masses and {\it WISE} derived mid-IR luminosities) are closely consistent with the Meidt et al. (2014) prediction of $M_*/L_{W1} = 0.60$. (Recall that both T11 and Meidt et al. use the same IMF and underlying BC03 models).


\begin{figure}
\includegraphics[width=\linewidth]{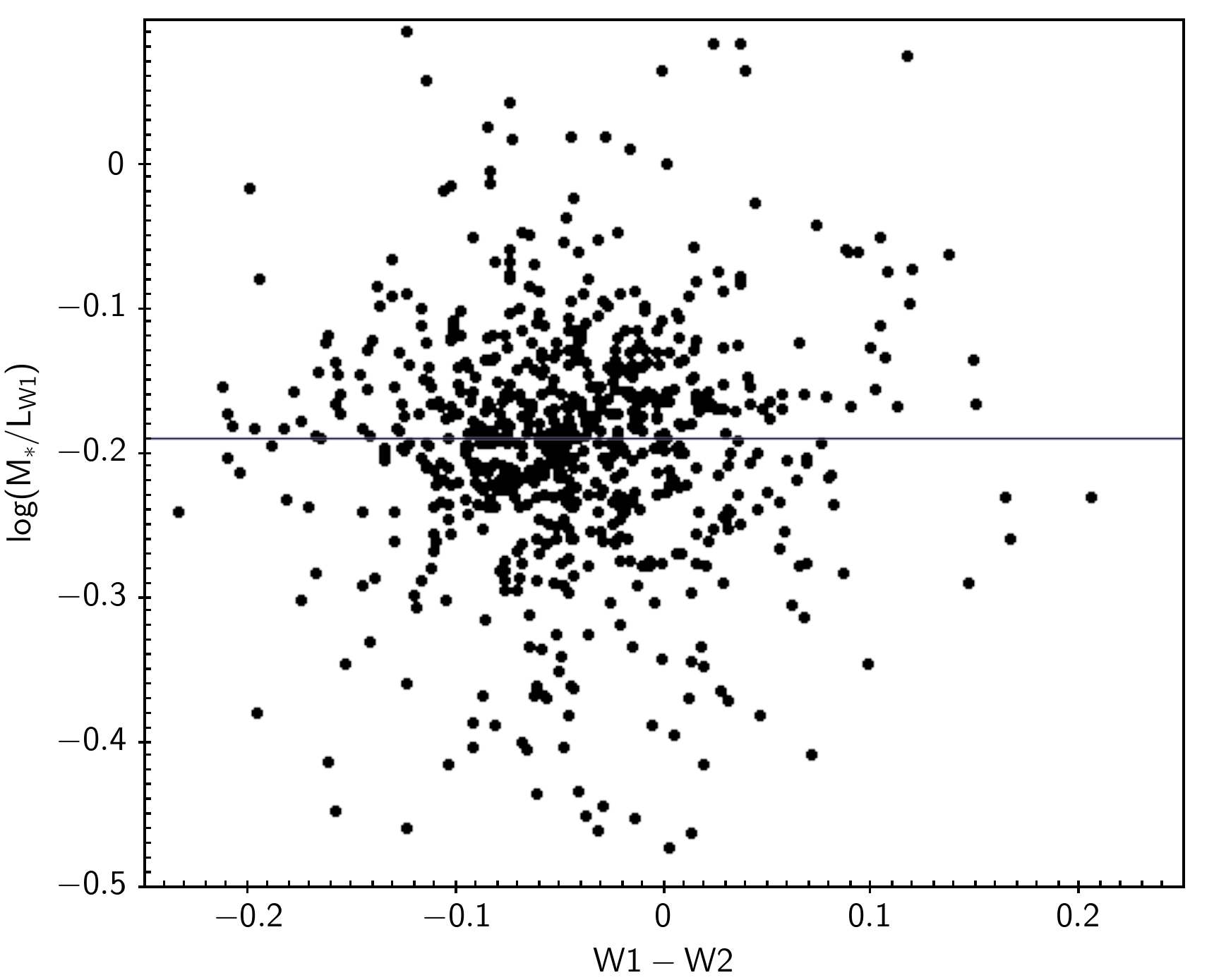}
\includegraphics[width=\linewidth]{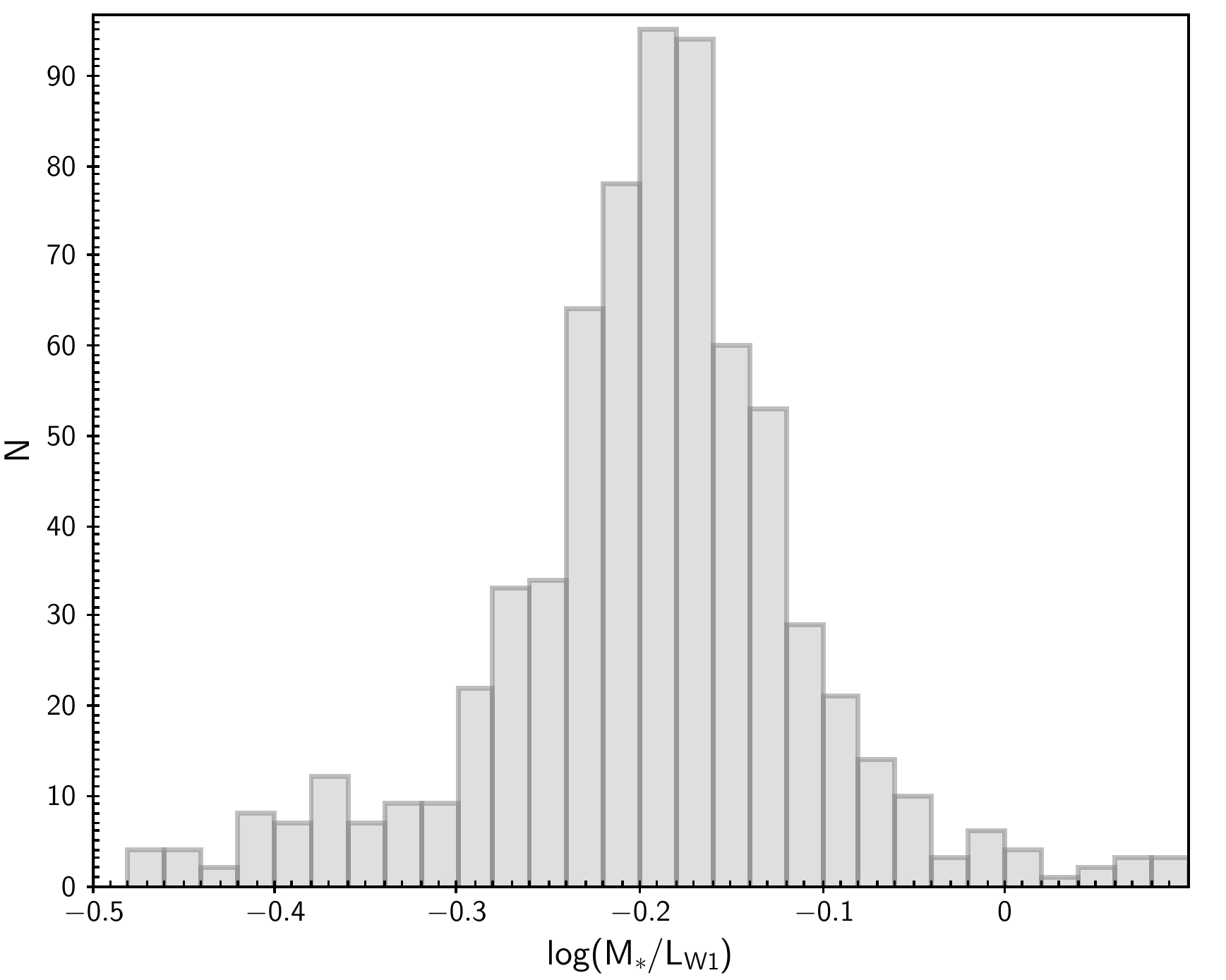}
\caption{(Upper panel) The stellar mass-to-light ratio (in solar units) in the $W1$ band versus intrinsic (redshift corrected) stellar population ($W1-W2$) colour for our sample galaxies. The horizontal line shows the mean value of the mass-to-light ratio. (Lower panel) Histogram of the mass-to-light ratios for our passive sample.
}
\label{ml-wise}
\end{figure}

As regards the spread, we can see that the standard deviation in the observed values is even smaller than might have been anticipated from the models discussed above. However, our sample is very tightly constrained to be passive so, for instance, we do not expect to be including young galaxies. We can check this from the GAMA derived ages (T11) which indeed show a narrow range around 6 to 7.5 Gyr. (There are very few GAMA galaxies, in general, with fitted ages above 9 Gyr, roughly the time betweeen $z=2.5$ and $z=0.15$). In addition, the SED fits also imply a narrow range in $Z$, around 0.4 to 1 $Z_{\odot}$. The stellar masses are all above $2 \times 10^9 M_{\odot}$, so we have no low metallicity dwarfs. 

For comparison, we can, as an example,
restrict attention to, say, solar metallicity models with ages 5 - 7 Gyr. From Norris et al. (2014) we would then expect a variation of only about 0.10 dex, as observed. Indeed, their predicted range in log($M_*/L_{W1})$ of $-0.22$ 
to $-0.11$ is also in excellent agreement with the values of log($M_*/L_{W1}$) seen for our sample, spanning almost exactly our $\pm 1 \sigma$ range $-0.24$ to $-0.14$. Lower metallicity ($0.1 Z_{\odot}$) models, while showing a similar sized spread, are systematically offset 
towards higher log($M_*/L_{W1}) = -0.14$ to $-0.06$, thus outside our $\pm 1 \sigma$ range.  

Equivalent results arise from the Meidt et al. (2014) modelling, which they present in terms of the star formation decay time $\tau$. Our GAMA-{\it WISE} sample has $\tau$ values largely between 0.8 and 3~Gyr from the T11 fits (i.e. stars formed rapidly, as expected) and Meidt et al.'s two most rapidly declining models give log($M_*/L_{W1}$) values between -0.21 and -0.13 for solar metallicity, again close to the range we obtain. They are about 0.06 higher for $Z= 0.2 Z_{\odot}$.

Thus we see that for most of our passive (and reasonably high metallicity) galaxies, choosing a fixed $M_*/L_{W1}$ of 0.65 and using the $W1$ luminosity, we can replicate the optical SED-derived $M_*$ to within a factor of about 1.1-1.2 and that the range of values we determine observationally for $M_*/L_{W1}$ is consistent with theoretical predictions.
 
We show the correspondance of the masses directly in Figure 2, where we plot the T11 masses (in $M_{\odot}$) against the masses derived from the $W1$ band luminosity simply via $0.65 (L_{W1}/L_{W1,\odot})$. This shows that as expected from the narrow $M_*/L_{W1}$ range in Figure 1, the masses agree to typically $\pm0.05$~dex. Further, the linear relationship shows that there are no significant mass dependent differences between the two - entirely observationally independent - approaches (though there may be a tentative suggestion that among the most massive galaxies the optical SED derived masses are more often slightly above, as opposed to below, the {\it WISE} derived masses).

\begin{figure}
\includegraphics[width=\linewidth]{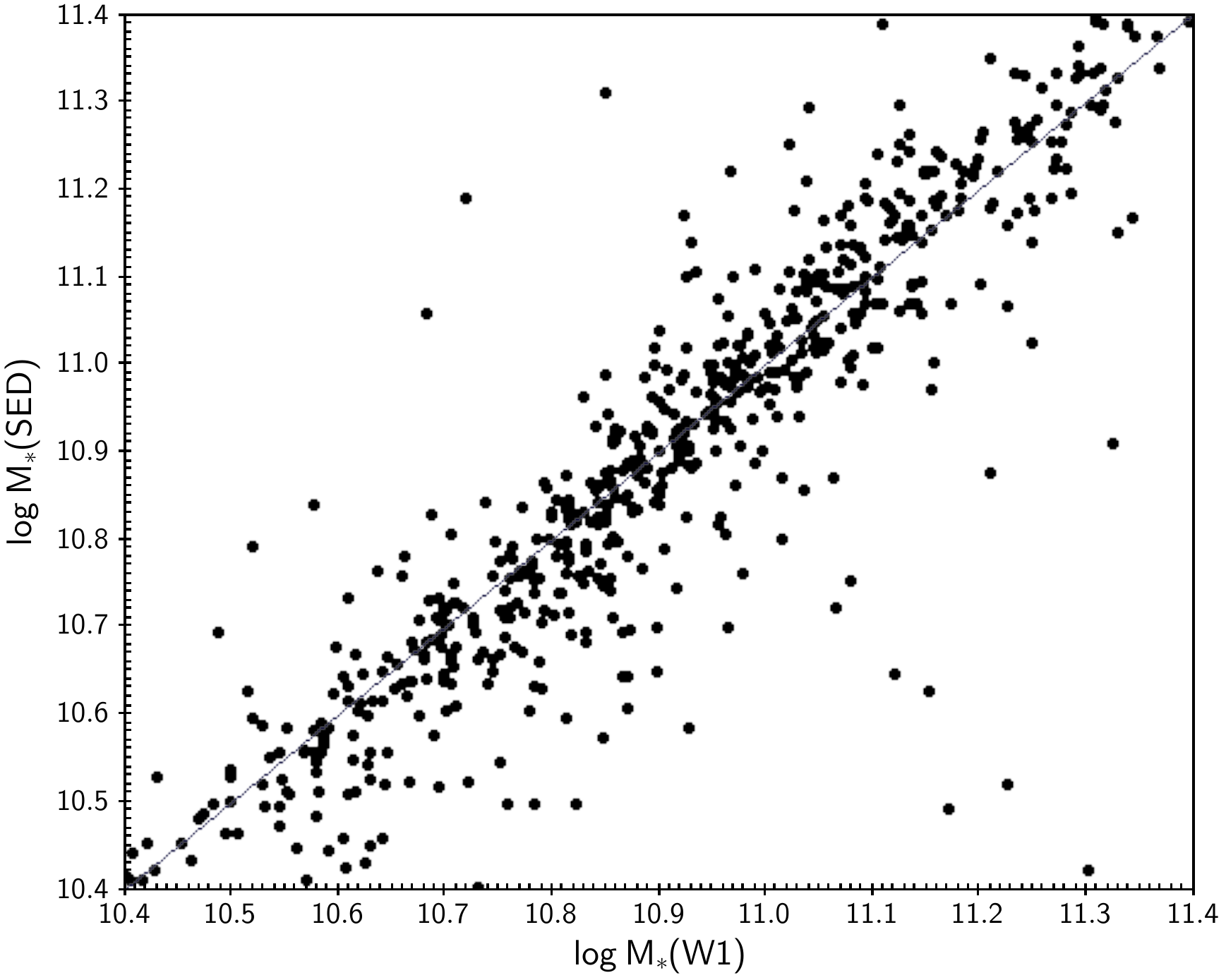}
\caption{Direct comparison of the optical SED-determined masses (from T11) and WISE-determined masses (see text) for our passive galaxy sample, both in units of $M_{\odot}$. The straight line shows the line of equal masses by each method.
}
\label{mass-from-wise}
\end{figure}

Returning to the scatter in more detail, we can compare the overall $\sigma \simeq 12$\% (i.e. $M_*/L_{W1} = 0.65 \pm 0.07$) to the expected errors. Errors in the calculated $L_{W1}$ luminosity will reflect the reported accuracy of the {\it WISE} magnitudes of $\simeq 0.03$ magnitudes (i.e. 3\% in flux) and the uncertainty due to our k-correction and the shape of the mid-IR SED (of order 5\% as discussed above). The galaxy-to-galaxy uncertainty in $M_*$ (either purely random or what T11 call `differential systematic errors', that is, those depending on age, SFH etc.) should then be at most 10\%, even if $M_*/L_{W1}$ has zero intrinsic scatter (for our particular objects). 

This is considerably better than might have been expected; Taylor et al (2010), for example, were only able to show that the errors in masses derived for SDSS galaxies were definitely less than 40\%. In particular, the small scatter in $M_*/L_{W1}$ implies that errors from the dust corrections in the SED fitting must be small, since they influence the SED-derived mass but not the $W1$ luminosity (see also Wright et al. 2017). Of course, we have a very specific sample of galaxies where differences in stellar populations are reduced and dust effects are small, typically a fitted E(B-V) of $0.1 \pm 0.1$. Nevertheless, it would seem that, at least for passive galaxies, T11 may have been unduly cautious about the  galaxy-to-galaxy uncertainties in their masses, though our analysis says nothing about the, likely much larger, global systematic uncertainties due to choice of SPS model, particularly the IMF (T11, Gallazzi \& Bell 2009, Conroy \& Gunn 2009, Gunawardhana et al. 2011, R\"ock et al 2015, Wright et al. 2017). We return to the question of systematic errors in Section 4.

The agreement between GAMA optical SED derived and mid-IR derived masses (from a simple constant $M_*/L_{W1}$) seen in Figure 2 is also better than might be inferred from some previous work, e.g., Poudel et al. (2016) who used MAGPHYS (da Cunha, Charlot \& Elbaz 2008) to fit SEDS across the whole range FUV to mid-IR. They found that they reproduced T11 masses only with $\sigma($log~$M_*) = 0.477$, i.e. a factor 3. This may be due to the typically large mismatch between the observed and model predicted fluxes in the {\it WISE} bands apparent from their overall fits (see their Figure 2).\footnote{Note that this problem does not occur with the GAMA and {\it WISE} data: Wright et al. (2017) show that the masses derived from MAGPHYS fits across a very large wavelength range (from Wright et al. 2016) are in very good agreement with those derived by the T11 method.} This underscores the value of our complementary method of testing the modelling of separate wavelength regimes.

\subsection{Population Parameters}

In terms of the SPS models, we would expect  $M_*/L_{W1}$ to increase both with increasing age and (slightly) with decreasing metallicity.

As noted above, our sample has quite a limited range in age (i.e. time since the onset of star formation), with a peak at $t =$ 7~Gyr and a steady decline to younger ages. Note that at these moderately old ages, differences in SEDs are also  produced in the T11 models by changes in the star formation timescale $\tau$, the formation epoch and the e-folding time both contributing to the stellar population's mean luminosity weighted age $t_*$.

 Figure 3 demonstrates that despite the limited range there is indeed the expected trend for higher $M_*/L_{W1}$ at older ages. Galaxies with ages above 7~Gyr have a mean log($M_*/L_{W1}) = -0.177 \pm 0.005$ while those below $t= 6.5$~Gyr have log($M_*/L_{W1}) = -0.232 \pm 0.007$. Given that uncertainties in the ages will shuffle galaxies across the borders, this is in reasonable agreement with a change from $-0.12$ to $-0.22$  for the solar metallicity Norris et al. (2014) models between the ages of 7 and 5~Gyr.  

The Meidt et al. (2014) models with short e-folding times (star formation completed quickly) give very similar predictions covering a range from $-0.13$ to $-0.21$ for both $Z = Z_{\odot}$ and $0.4 Z_{\odot}$. With our data, we see corresponding variations to those in Figure 3 if we use either $\tau$ or $t_*$ in place of the age $t$. In particular for the fall-off time $\tau$, we see the expected small shift to higher log($M_*/L_{W1}$) with decreasing $\tau$ (increasing mean population age), again from $\simeq -0.23$ to $-0.18$, similar to the Meidt et al. models.

\begin{figure}
\includegraphics[width=\linewidth]{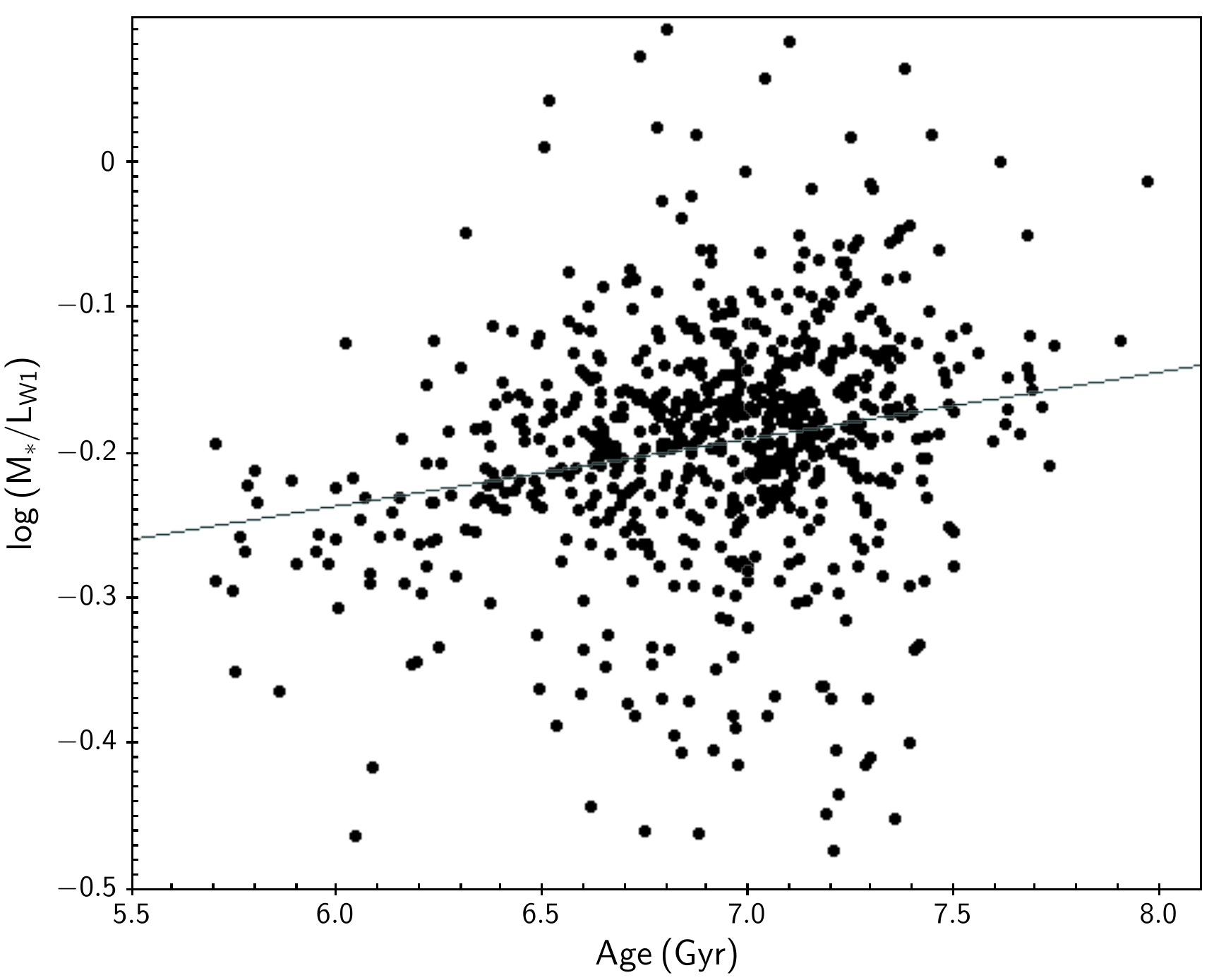}
\caption{The stellar mass-to-light ratio in the $W1$ band versus galaxy age $t$ as derived from the T11 optical SED fitting. The straight line shows the overall trend for increasing mass-to-light ratio with age as described in the text. 
}
\label{ml-wise-age}
\end{figure}

Notice that, {\em inter alia}, these results not only give support for the SPS modelling across the optical and {\it WISE} bands but also, at least in a global sense, for the validity of the age determinations (from similar models but different data). Likewise, the failure of very old ($\geq 10$~Gyr) models to predict the correct $M_*/L_{W1}$ (they lie well above the data) is consistent with the SED modelling generating very few objects with such ages even among our passive galaxies.

The age-related variation of $M_*/L_{W1}$ which we see in our data will, of course, contribute to the spread seen in the data. In principle it could be used to further sharpen the expected value of $M_*/L_{W1}$ for any particular galaxy, as noted previously by Meidt et al., though in practice it is clearly a small effect amongst our old, passive galaxies. If we choose an age dependent $M_*/L_{W1}$, varying by the amount expected from the models and seen empirically in Figure 3, we do obtain a slight reduction in the scatter in the ratio of $\: M_*$ from T11 to that derived from $L_{W1}$, as in Figure 2, but by only $\simeq 0.003$~dex. In any case, while offering further support to the modelling, this is not useful in the practical context of using the WISE data to determine $M_*$, since the ages are only available from SED fitting or other more detailed measurements. 

Moving on to metallicities, as noted earlier, differences in $M_*/L_{W1}$ between galaxies of similar age with $Z$ varying between a few tenths and 1 $Z_{\odot}$ are expected to be very small, $\sim 0.01$~dex. Within our data, and given the difficulty in meaningfully constraining $Z$ from the optical-NIR SED, unsurprisingly we see no significant effect. Splitting at the median $Z$ we find a mean log($M_*/L_{W1}$) of $-0.189 \pm 0.005$ at higher $Z$ and $-0.199 \pm 0.005$ for lower $Z$. 

Finally, it is also interesting to briefly return to the joint $M_*/L_{W1}$ - colour plot of Figure 1.
At fixed age, Norris et al (2014) find that changing the metallicity even between 0.1 and 1 $Z_{\odot}$  induces a change in model ($W1-W2$) colour of merely 0.04 magnitudes (e.g. from $-0.04$ to $-0.08$ at 7~Gyr). The larger potential range in colour presented by Norris et al. is only obtained by including very low metallicities, not found among large early type galaxies. Indeed our sample has such a tightly constrained metallicity range that, from Meidt et al., expected colour variations for 0.4 to 1~$Z_{\odot}$ should only be at the 0.01 magnitudes level.  Further, the predicted change in $(W1-W2)$ with age at fixed $Z$ is minimal, even less than 0.01 magnitudes.  R\"ock et al. (2015) find colour differences at the same levels for various different SSP models. 

These ranges are much smaller than the expected errors from the {\it WISE} photometry of around 0.056 magnitudes which completely dominate the $1 \sigma$ observed colour spread of 0.063 magnitudes (about a mean of $-0.044$ for the redshift corrected colour). Note that there are a number of outliers at apparently impossible colours relative to the models of metal rich populations (e.g. around 0.05 to 0.1), but these could be merely the (2-3~$\sigma$) outliers in the error distribution. The colour errors are clearly too large to see any correlation between $Z$ (from T11) and $(W1-W2)$ in the present data.

 The actual colours are in reasonable agreement between the models and the data, though the observed colour range is slightly too red, on average, given the restricted passive sample we use. For instance, as above, Norris et al. obtain ($W1-W2) = -0.08$ for their solar abundance models with ages similar to our objects, compared to our mean of $-0.044$. R\"{o}ck et al. (2015) use a variety of models but again for solar or slightly sub-solar abundances, find that their expected {\it WISE} colours are always close to $-0.08$ for reasonably old populations (their figure 11). However, regardless of whether this is due to a slight calibration issue in either the data or the models, even if we change the $W1$ magnitudes by the whole 0.04 shift, this changes the corresponding luminosities (and hence deduced $M_*/L_{W1}$) by only 4\%, well within the errors discussed earlier.

\subsection{Star Forming Galaxies}
For completeness, we can note that if we look at the whole low $z$ GAMA-{\it WISE} sample, rather than our specific passive sample, then the star forming galaxies extend the mass-to-light ratios to lower values (typically log($M_*/L_{W1}) \sim -0.4 \pm 0.2$), alongside redder $(W1-W2)$ and $(W2-W3$) colours, as already shown and discussed in Cluver et al. (2014; see particularly their figure 7). Querejeta et al (2015) have presented a method, based on the mid-IR colours, for separating the mid-IR emission into components from the old stellar population and from the dust associated with star formation, the former then being usable to determine stellar masses even in the presence of the latter.

\section{Discussion}

The very positive outcome of this work is that we confirm model predictions that a fixed mid-IR stellar mass-to-light ratio for passive galaxies can replicate masses from optical SED fits to an uncertainty level of around 12\%. This implies  that the GAMA catalogued masses for such galaxies are good to better than 10\%, i.e. better than the accuracy cautiously claimed in T11 for galaxies in general. Looked at from the point of view of our original question, it also implies that the same SPS models (in our case from BC03) give consistent results when applied to two separate wavelength regimes, at least for old stellar populations. Essentially, one can use the optical-NIR SED to `predict' the $W1$ flux to a surprisingly good level of accuracy (c.f. Blanton \& Roweis 2007 for the case of NIR data). Put another way, this argues that the SED based stellar mass estimates of passive galaxies are very good, but on the other hand that they are not actually necessary, in that $W1$ luminosity alone does an equally good job.

Of course, the actual masses will be dependent on the assumed IMF, Meidt et al. (2014) calculating that their  standard $M_*/L_{W1} = 0.60$ for a Chabrier IMF would be increased to 1.06 for a Salpeter IMF, for example. However, increases in $M/L$ will be general across the optical and IR regimes (increased numbers of low mass stars generating little of the light at any of these wavelengths), so does not materially affect our mass comparisons. In addition, the fact that we have limited age and mass ranges for our galaxies will also negate the effect from any possible epoch or mass dependent variations in the IMF (e.g. van Dokkum 2008 and La Barbera et al. 2013, respectively), so these will not introduce any extra scatter. 
 
We should mention one caveat here, though, the possibility of compensating errors reducing the scatter in the mass-to-light ratios (both here and in T11's work using $(g-i$) colours). 
The derived $M_*$ values are formally independent of the {\it WISE} photometry, so any systematic errors in $M_*$ are not tied directly to mid-IR colours. However, any systematics that depend on stellar population properties might show up indirectly as a function of {\it WISE} colours (to the extent that {\it WISE} colours trace stellar populations). It might be conceivable, for instance, that there is a systematic error in $M_*$ values that is a function of the stellar population (i.e. SED) such that it either amplifies {\it or} diminishes any true variation in $M_*/L_{W1}$ when looked at as a function of $(W1-W2)$. This may be the ultimately limiting factor in the analysis of the stellar population fits, but further investigation of this point is outside the scope of the present paper.

Both the mean and range of $M_*/L_{W1}$ are consistent between the joint data (GAMA mass and {\it WISE} luminosity) and the models.
In addition, we see the predicted small change of $M_*/L_{W1}$ with age, but not the even smaller effect with metallicity, if we use the T11 stellar population fits for these parameters. This also implies that the derived ages, at least, are genuinely physically meaningful, even though the ranges in these parameters are small in our particularly passive galaxy sample. We do not see the predicted variation of $(W1-W2)$ colour with metallicity, but this is expected to be very small across the range we sample and is currently drowned by the errors.

From a broader viewpoint, our results give added credence to the SPS modelling of the optical SED (e.g. T11) and the mid-IR SED ({\it WISE}) providing  consistent methods for determining stellar masses. The observations of the $W1$ band mass-to-light ratio for old passive galaxies also give us confidence that the modelled variations of $M_*/L_{W1}$ with age - and hence observables such as the intrinsic ($u-r$) colour - are correct, giving us a direct method of determining stellar masses across a wider range of stellar populations than studied here. Furthermore, if the current models can indeed be used for galaxies at younger ages then observing at longer infra-red wavelengths with e.g. {\it Spitzer} or, in the future MIRI on JWST, should give reliable stellar masses for high redshift galaxies. For instance, the 10 and 12.8~$\mu$m MIRI imaging filters are close to rest-frame $W1$ and $W2$ for a galaxy at $z=1.8$, when the age of the Universe was $\simeq 3.7$~Gyr. The models suggest a smooth variation of $M_*/L_{W1}$ with age, R\"{o}ck et al. (2015), for instance, showing that for SSPs (i.e. instantaneous bursts) at ages 1 - 3~Gyr, $M_*/L_{W1}$ should vary from 0.2 to 0.4, compared to the 0.6 for the older galaxies discussed in the present paper. $M_*/L_{W1}$ will also be affected by ongoing star formation, but from the models the values are likely to be similar, Meidt et al. (2014) obtaining around 0.45 for their solar metallicity models with very long e-folding times (so effectively constant star formation), albeit at the present day, in agreement with the observations of star forming galaxies reported in Cluver et al. (2014).

\section*{Acknowledgements}
We thank the referee for useful comments which helped us clarify some points. GAMA is a joint European-Australasian project based around a spectroscopic campaign using the Anglo-Australian Telescope. The GAMA input catalogue is based on data taken from the Sloan Digital Sky Survey and the UKIRT Deep Sky Survey. Complementary imaging of the GAMA regions is being obtained by a number of independent survey programmes including GALEX MIS, VST KiDS, VISTA VIKING, {\it WISE}, Herschel-ATLAS, GMRT and ASKAP providing UV to radio coverage. GAMA is funded by the STFC (UK), the ARC (Australia), the AAO and participating organisations. The GAMA website is http://www.gama-survey.org/. This work has made considerable use of TOPCAT, written and maintained by Mark Taylor via a STFC grant at the University of Bristol.

\section*{References}
{\small 
Alam S., Albareti F.D., Allende Prieto C. et al., 2015, ApJS,
 
219, 12\\ 
Baldry I.K., Driver S.P., Loveday J. et al., 2012, MNRAS, 
421, 

621\\
Bell E.F, de Jong R.S., 2001, Apj, 550, 212\\
Bell E.F., McIntosh D.H., Katz N., Weinberg M.D., 2003, ApJS,

 149, 289\\
Blanton M.R., Roweis S., 2007, AJ, 133, 734\\
Bressan A., Marigo P., Girardi L., Salasnich B., Dal Cero C., 

Rubele S., Nanni A., 2012, MNRASA, 427, 127\\
Brown M.J.I, Moustakas J., Smith J.-D.T. et al., 2014, ApJS, 

212, 18\\
Brough S., van de Sande J., Owers M.S. et al., 2017, 
arXiv

170401169\\
Bruzual G., Charlot S., 2003, MNRAS, 344, 1000\\
Calzetti D., Armus L., Bohlin R.C., Kinney A.L., Koornneef J., 

Storchi-Bergmann T., 2000, ApJ, 533, 682\\
Chabrier G., 2003, PASP, 115, 763\\
Chang Y.-Y., van der Wel A., da Cunha E., Rix H.-W., 2015,

ApJS, 219, 8\\
Cluver M., Jarrett T.H., Hopkins A.M. et al., 2014, ApJ, 
782, 90\\
Colless M., Dalton G., Maddox S. et al., 2001, MNRAS, 
328, 1039\\
Conroy C., 2013, ARA\&A, 51, 393\\
da Cunha E., Charlot S., Elbaz, D., 2008, MNRAS, 388, 1595\\
Davies L.J.M., Driver S.P., Robotham A.S.G. et al., 2016, 

MNRAS, 461, 458\\
Driver S.P., Norberg P., Baldry I.K. et al., 2009, A\&G, 
50, 12\\
Driver S.P., Hill D.T., Kelvin L.S. et al., 2011, MNRAS, 
413, 971\\
Driver S.P., Wright A.H., Andrews S.K. et al., 2016, MNRAS,

455, 3911\\
Edge A., Sutherland W., Kuijken K. et al., 2013, Messenger,

154, 32\\
Gallazzi A., Bell E.F., 2009, ApJS, 185, 253\\
Gallazzi A., Charlot S., Brinkmann J., White S.D.M., Tremonti 

C., 2005, MNRAS, 362, 41\\
Gunawardhana M.L.P, Hopkins A.M., Sharp R.G. et al., 2011,

MNRAS, 415, 1647\\
Gunawardhana M.L.P., Hopkins A.M., Bland-Hawthorn J. et al.,

2013, MNRAS, 433, 2764\\
Hill D.T., Kelvin L.S., Driver S.P. et al., 2011, MNRAS,
412, 765\\
Hopkins, A.M., Driver S.P., Brough S. et al, 2013, MNRAS,
430, 

2047\\
Huang J.S., Ashby M.L.N., Barmby P. et al., 2007, ApJ,
664, 840\\
Jarrett T.H., Cohen M., Masci F. et al., 2011, ApJ, 735, 112\\
Jarrett T.H, Masci F., Tsai C.W. et al., 2013, AJ, 145, 6\\
Kauffmanm G., Heckman T.M., White S.D.M. et al., 2003a, 

MNRAS, 341, 33\\
Kauffmanm G., Heckman T.M., White S.D.M. et al., 2003b, 

MNRAS, 341, 54\\
Kelvin L.S., Driver S.P., Robotham A.S.G. et al., 2012,
MNRAS, 

421, 1007\\
La Barbera F., Ferreras I., Vazdekis A. et al., 2013,
MNRAS, 

433, 3017\\
Liske J., Baldry I.K., Driver S.P. et al., 2015, MNRAS,
452, 2087\\
Maraston C., 2005, MNRAS, 362, 799\\
Martinsson T.P.K., Verheijen M.A.W., Westfall K.B., 
Bershady

M.A., Andersen D.R., Swaters R.A., 2013, A\&A, 
557, 131\\ 
McGaugh S.M., Schombert J.M., 2014, AJ, 148, 77\\
Meidt S.E., Schinnerer E., van de Ven G. et al., 2014, ApJ, 
788, 

144\\
Mendel J.T., Simard L., Palmer M., Ellison S.L., Patton D.R., 

2014, ApJS, 210, 3\\
Moster B.P., Sommerville R.S., Maulbetsch C., van den Bosch 

F.C., Maccio A.V., Naab, T., Oser L., 2010, ApJ, 710, 903\\
Neill J.D., Seibert M., Tully R.B. et al., 2014, 792, 129\\
Noeske K.G., Weiner B.J., Faber S.M. et al., 2007, ApJL, 
660, 

L43\\
Norris M.A., Meidt S., van de Ven G., Schinnerer, E, 
Groves B., 

Querejeta M., 2014, ApJ, 797, 55\\
Norris M.A., Van de Ven, G., Schinnerer E. et al., 2016, ApJ, 

832, 198\\
Peletier R., Kutdemir E., van der Wolk G. et al., 2012, MNRAS, 

419, 2031\\
Poudel A., Hein\"{a}m\"{a}ki, Nurmi P., Teerikorpi P., Tempel E., 

Lietzen H., Einasto M., 2016, A\&A, 590, 29\\
Querejeta M., Meidt S.E., Schinnerer E. et al., 2015, ApJS, 219, 

5\\
R\"{o}ck B., Vazdekis A., Peletier R.F., Knapen J.H., 
Falc\'{o}-Barroso 

J., 2015, MNRAS, 449, 2853\\
Spitler L.R., Forbes D.A., Beasley M.A., 2008, MNRAS, 389, 1150\\
Stern D., Assef R.J., Benford D.J. et al., 2012, ApJ, 753, 30\\
Taylor E.N., Franx M., Brinchmann J., van der Wel A., van 

Dokkum P.G., 
2010, ApJ, 722, 1\\
Taylor E.N., Hopkins A.M., Baldry I.K. et al., 2011, MNRAS, 

418, 1587\\
Taylor M.B., 2005, Astronomical Data Analysis Software and 

Systems XIV, ASPC 347, 29\\
Thanjavur K., Simard L., Bluck A.F.L., Mendel T., 2016, 

MNRAS, 459, 44\\
Tojeiro R., Heavens A.F., Jimenez R., Panter B., 2007, MNRAS, 

381, 1252\\
Tremonti C.A., Heckman T.M., Kauffmann G. et al., 2004, ApJ, 

613, 898\\
van Dokkum P.G., 2008, ApJ, 674, 29\\
Walcher J., Groves B., Budavari T., Dale D., 2011, Ap\&SS, 
331,~1\\
Wen X.Q., Wu H., Zhu Y.-N., Lam M.I., Wu C.-J., Wicker J., 

Zhao Y.-H., 2013, MNRAS, 433, 2946\\
Wright A.H., Robotham A.S.G., Bourne N. et al., 2016, MNRAS,

460, 765\\
Wright A.H., Robotham A.S.G., Driver S.P. et al., 2017, arXiv

170504074\\
Yan L., Donoso E., Tsai C.-W. et al., 2013, AJ, 145, 55\\
York D.G., Adelman J., Anderson J.E. et al, 2000, AJ, 
120, 1579\\
Zibetti S., Charlot S., Rix H.-W., 2009, MNRAS, 400, 1181
}
\end{document}